\documentstyle[aps,psfig,twocolumn]{revtex}

\begin{document}

%\twocolumn
\date{17 June 1999}
%\preprint{}
%\draft

\title{Multiple quantum phases in artificial double-dot molecules}

\author{Massimo Rontani$^{1,2}$, F.~Rossi$^{1,2,3}$, F.~Manghi$^{1,2}$, 
and E.~Molinari$^{1,2}$}
\address{$^1$ Istituto Nazionale per la Fisica della Materia (INFM)}
\address{$^2$ Dipartimento di Fisica, Universit\`a di Modena
e Reggio Emilia, Via Campi 213/A, I-41100 Modena, Italy}
\address{$^3$ Dipartimento di Fisica, Politecnico di Torino,
Corso Duca degli Abruzzi 24, I-10129 Torino, Italy}

\maketitle

\begin{abstract} %\baselineskip=2.5ex

We study coupled semiconductor quantum dots theoretically
through a generalized Hubbard approach, where intra- and inter-dot
Coulomb correlation, as well as tunneling effects
are described on the basis of realistic electron 
wavefunctions. We find that the ground-state configuration
of vertically-coupled double dots undergoes non-trivial quantum transitions
as a function of the inter-dot distance $d$; at intermediate
values of $d$ we predict a new phase that should be observable
in the addition spectra and in the magnetization changes.

\end{abstract}

\pacs{73.20.Dx,  73.61.-r, 85.30.Vw}

%\begin{multicols}{2}
\narrowtext

Semiconductor quantum dots (QDs) are nano- or mesoscopic structures
that can be regarded as `artificial atoms' because of the
three-dimensional carrier confinement and the resulting discrete
energy spectrum. They are currently receiving great attention
because they can be designed to study and exploit new physical phenomena:
On the one hand, the nature and scale of electronic confinement
allows the exploration of regimes that are not accessible in conventional
atomic physics; on the other hand, they lead to novel devices
dominated by single- or few-electron effects \cite{reviews,hawrylak}.

One of the important challenges at this point is to understand the
fundamental properties of coupled quantum dots, the simplest structures
that display the interactions controlling potential quantum-computing
devices. Here also the interdot
coupling can be tuned through external parameters, far out
of the regimes known in `natural molecules' where the ground-state
interatomic distance is dictated by the nature of bonding.
We expect that new phenomena will occur in these `artificial molecules'
(AMs)
when the relative importance of Coulomb interaction and single-particle
tunneling is varied. Through the study of such phenomena,
coupled dots may become a unique laboratory to explore electronic
correlations.

In this paper we analyze ground and
excited few-particle states of realistic double quantum dots (DQDs)
using a theoretical scheme
that fully takes into account intra- and inter-dot
many-body interactions. We focus on strongly confined 
vertically-coupled DQDs,
and show that, for a given number of electrons, $N$,
the ground state configuration undergoes non-trivial
quantum transitions as a function of the inter-dot distance $d$;
we identify specific ranges of the DQD parameters characterizing
a `coherent' molecular phase, where tunneling effects dominate,
and a phase where the inter-dot interaction is purely electrostatic.
At intermediate values of $d$ we predict a new phase that should
be observable by means of transport experiments
in the addition spectra and in the magnetization changes.

From the theoretical point of view, a major difficulty is
that we cannot make use of perturbative schemes \cite{perturb}
in the calculation of the DQD many-body ground state,
since we want to investigate all inter-dot coupling regimes.
We thus write the many-body hamiltonian $H$ within a
generalized Hubbard (GHH) approach \cite{theory},
and chose a basis set formed by suitable single-particle
wavefunctions localized on either dot ($i=1,2$), and characterized by
orbital
quantum numbers, $\alpha$ (to be specified below),
and by the spin quantum number, $\sigma$.
On this basis the many-body hamiltonian is
\begin{eqnarray}
\hat{H}&=&\sum_{i\alpha\sigma}\tilde{\varepsilon}_{\alpha}
\hat{n}_{i\alpha\sigma}
-t\sum_{\alpha\sigma\left< ij \right>}
{\hat{c}_{i\alpha\sigma}}^{\dagger}
\hat{c}_{j\alpha\sigma}
+\frac{1}{2}\sum_{i\alpha\beta\sigma}U_{\alpha\beta}
\hat{n}_{i\alpha\sigma}\hat{n}_{i\beta-\sigma}  \nonumber \\
&& +\frac{1}{2}\sum_{i\alpha\beta\sigma}
\left(U_{\alpha\beta}-J_{\alpha\beta} \right)
\hat{n}_{i\alpha\sigma}\hat{n}_{i\beta\sigma}
+\sum_{\alpha\beta\sigma\sigma^{\prime}}
\tilde{U}_{\alpha\beta}\hat{n}_{1\alpha\sigma}
\hat{n}_{2\beta\sigma^{\prime}}. \nonumber
\end{eqnarray}
The first two addenda are the
single-particle on-site and hopping term respectively:
$\tilde{\varepsilon}_{\alpha}$ are the single-particle energies,
$t$ the tunneling parameter; $\hat{n}$ are the occupation numbers
and $\hat{c}$ ($\hat{c^{\dagger}}$) the creation (destruction) operators.
The third and fourth terms account for intra-dot Coulomb interaction 
between electrons with
antiparallel and parallel spins, respectively:
$U$ and $J$ are the intra-dot Coulomb and exchange integrals.
Finally, the last term represents the inter-dot Coulomb coupling,
$\tilde{U}$ being the inter-dot Coulomb integrals.
These integrals are in turn expressed in terms
of single-particle states:
\begin{displaymath}
U_{\alpha\beta}=\int
\frac{ {e^2 \left|{\phi}_{i\alpha}\!\left({\bf r}\right)\right|}^2
{\left|{\phi}_{i\beta}\!\left({\bf r^{\prime}}\right)\right|}^2}
{\kappa_r\left|{\bf r-r^{\prime}}\right|}
{\rm\,d}{\bf r}{\rm\,d}{\bf r^{\prime}}; \;\;
\end{displaymath}
\begin{displaymath}
\tilde{U}_{\alpha\beta}=\int
\frac{ {e^2\left|{\phi}_{1\alpha}\!\left({\bf r}\right)\right|}^2
{\left|{\phi}_{2\beta}\!\left({\bf r^{\prime}}\right)\right|}^2}
{\kappa_r\left|{\bf r-r^{\prime}}\right|}
{\rm\,d}{\bf r}{\rm\,d}{\bf r^{\prime}};
\end{displaymath}
\begin{displaymath}
J_{\alpha\beta}=\int
\frac{ {e^2{\phi}_{i\alpha}}^{\!*}\!\left( {\bf r} \right)
{{\phi}_{i\beta}}^{\!*}\!\left({\bf r^{\prime}}\right)
{\phi}_{i\alpha}\!\left({\bf r^{\prime}}\right)
{\phi}_{i\beta}\!\left({\bf r}\right) }{\kappa_r\left|{\bf r-r^{\prime}}
\right|}
{\rm\,d}{\bf r}{\rm\,d}{\bf r^{\prime}}\! ;
\end{displaymath}
with ${\phi}_{i\alpha}\!\left({\bf r}\right)$
full three-dimensional single-particle wavefunctions
(obtained within the usual envelope-function approximation),
$e$ electronic charge and $\kappa_r$ relative dielectric constant.

Here we shall consider a specific type of nanostructures, namely
vertically coupled cylindrical DQDs \cite{austing};
the ingredients entering the Hamiltonian can therefore
be defined explicitly.
For simplicity, $V(r)$ can be assumed to be separable in the $xy$ and $z$
components; the profile is taken to be 
parabolic in the $xy$ plane, with confinement energy $\hbar\omega$, 
and a symmetric double quantum well (QW) along $z$  \cite{austing,tarucha}.
The eigenstates of the $xy$ harmonic
potential are the usual Fock-Darwin states
$\left|\alpha\right> = \left|(n,m)\right>$ ($n$, $m$ radial and angular
quantum numbers)  \cite{hawrylak}.
Along $z$, the QW thickness is such that only the lowest eigenstates
(symmetric, $\left|s\right>$, and antisymmetric $\left|a\right>$) are
relevant for the low-energy spectrum.
From these, we construct a complete set of states that are localized
on either dot (see inset of Fig.~1):
$\left|1\right>=\left(\left|s\right>+
\left|a\right>\right)/\sqrt{2}$ and $\left|2\right>=
\left(\left|s\right>-\left|a\right>\right)/\sqrt{2}$.
The basis we use is therefore the direct product
$ \left|i\alpha\sigma\right>=\left|i\right>\otimes\left|\alpha\right>
\otimes\left|\sigma\right>$.
The single-particle energies are
$\tilde{\varepsilon}_{\alpha}=\varepsilon_{\alpha}+
(\epsilon_s+\epsilon_{a})/2$, and the tunneling parameter is
$t=(\epsilon_{a}-\epsilon_s)/2$,
with $\epsilon_s$, $\epsilon_{a}$ double-well eigenenergies,
and $\varepsilon_{\alpha}$ oscillator energies.

Using the above expressions, the three-dimensional Coulomb 
integrals and the tunneling parameter are computed directly
from the single-particle states for each sample. Fig.~1
shows the result for $t$, $U_{\alpha\beta}$ and
$\tilde{U}_{\alpha\beta}$ [$\alpha=\beta=(0,0)$] calculated
for a GaAs/AlAs DQD ($\kappa_r=$ 12.98, effective mass 0.065
$m_e$) with $\hbar \omega=$ 10 meV, as a function of the
interdot distance (barrier width) $d$.  Note that already
around $d=5$ nm the inter-dot Coulomb integral exceeds the
single-particle hopping parameter.

To obtain the many-body energies and eigenstates,
the Hamiltonian $\hat{H}$ is then diagonalized exactly
for each value of $N$ (total number of electrons)
on each configuration subspace labeled by the quantum numbers $S$
($z$-component of the total spin) and $M$ ($z$-component of the
total angular momentum).

This approach has two main advantages:
First, it allows us to solve the many-body problem
consistently in the different coupling regimes,
from the limit where $t$ dominates over the Coulomb integrals
to the opposite limit where $t$ is negligible.
Second, it can provide quantitative predictions for given
DQD structures, since it contains no free parameters and
uses realistic ingredients ($t$, ${U}_{\alpha\beta}$,
$\tilde{U}_{\alpha\beta}$, ${J}_{\alpha\beta}$) calculated
for each nanostructure \cite{foot1}.

Fig.~2 shows the calculated ground-state energies $E_N$ of
correlated $N$-particle states as a function of the inter-dot
distance $d$ \cite{notation}. The three lowest excited states are
also shown for comparison. As expected, when $d$ is large, the
system behaves as two isolated QDs. With decreasing $d$, some of
the many-body excited states, favoured by Coulomb interactions,
become lower in energy. For $N \leq 3$ a single quantum phase
transition occurs, below which the new ground state is a
molecular-like state. For $N>3$, two successive transitions take
place, at $d=d_b$ and $d=d_a$, and an intermediate non-trivial
phase is predicted to occur in the range $d_a<d<d_b$. Note that
this phase is stable in a relatively large range of $d$ values,
which depends on the number of electrons. An accurate determination
of $d_a$ and $d_b$ requires the correct inclusion of all inter-dot
coupling terms, including the inter-dot Coulomb integrals. If the
latter are neglected, all quantum transitions occur for smaller
values of the inter-dot distance, and the $d$-range of the
intermediate phase is underestimated significantly; had also the
other many body terms been  neglected, the intermediate phase would
disappear leaving  a simple crossover  from a  molecular to an
atomic like regime at $d=da=db$.

To understand the nature of the different phases it is useful to
examine the many-body states in terms of the single Slater
determinants that contribute to each of them. Here we discuss
explicitly the 4-electron case with the help of the insets in
Fig.~2, but the same reasoning can be followed for the other cases.
Both in the case of very small  and very large  interdot distances
the ground state can  be essentially described in terms of a single
Slater determinant: for large values of $d$ ($d>d_b$), the relevant
configurations for the ground state have two electrons in the
lowest level of each isolated dot; the $\left|s\right>$ and
$\left|a\right>$ extended `molecular' orbitals derived from the
lowest `atomic' states are of course almost degenerate and both
filled with two electrons. In  the opposite limit  ($d<d_a$), by
expanding the localized atomic orbitals in terms of molecular
orbitals we recognize that the $\left|s\right>$ state derived from
the lowest atomic state is filled with two electrons, while the
corresponding $\left|a\right>$ molecular state is empty. The two
remaining electrons occupy the next bonding molecular orbitals
---derived from the higher $p_x$ and $p_y$ levels--- with parallel
spin, in such a way that $S$ is maximized. This is the
manifestation of Coulomb interaction which leads to Hund's rule for
molecules. 

In these two extreme phases the single particle picture is
essentially correct, provided that the appropriate basis set
(either localized or extended  orbitals) is used.  In the
intermediate phase, $d_a<d<d_b$,  this is no longer true and the
ground state is a mixture of different Slater determinants in
any basis set \cite{Eto}.  In this sense again the intermediate
phase exhibits an intrinsic many-body character.  Coulomb direct
and exchange terms,  responsible for the selection of the global
quantum numbers,  determine a new ground state configuration, where
both  $S$ {\it and} $M$ are maximized.

We obtain a clear evidence of the different electronic distribution
in the three quantum phases by calculating the spin-dependent
electronic pair correlation function, defined as
%%%%%
$$g_{\sigma',\sigma''}({\bf \rho},z',z'')=\int d{\bf R}
\left <\hat\Psi^{\dagger}_{\sigma'}({\bf r'})
\hat\Psi^{\dagger}_{\sigma''}({\bf r''})
\hat\Psi_{\sigma''} ({\bf r''})
\hat\Psi_{\sigma'} ({\bf r'}) \right >,$$
where ${\bf R}=({\bf \rho}'+{\bf \rho}'')/2$ and
${\bf \rho}=({\bf \rho}'-{\bf \rho}'')$.
Here ${\bf \rho}' = (x', y')$, ${\bf \rho}'' = (x'', y'')$ are the
in-plane spatial coordinates of the electrons in the pair, and
$z'$ and $z''$ are their coordinates along $z$, that will be
kept fixed at the center of either QD, i.e. in
${z_1}$ or ${z_2}$. 
${\bf r}'=({\bf \rho}',z')$, ${\bf r}''=({\bf \rho}'',z'')$.
$\sigma'$, $\sigma''$ are the spin variables and can assume the values
$\uparrow$ or $\downarrow$.
In Fig.~3 we plot for example
$g_{\downarrow,\downarrow}(x,y,z',z'')$
for a double QD with $N=6$ electrons,
at three values of the inter-dot distance corresponding
to different quantum phases. Here $\downarrow$
represents the minority spins.
For  $d<d_a$ (left column), it is indeed apparent that the pair
correlation
function is the same
when both electrons are in the same dot [$z'=z''={z_1}$, panel (a)]
or on different dots [$z'={z_1}$ and $z''={z_2}$, panel (b)].
In this sense, the system behaves `coherently'. For $d>d_b$ (right
column),
the maps indicate that the probability of finding two
$\downarrow$-electrons in the same dot is negligible,
consistently with the picture of isolated dots. 
In the intermediate phase
$d_a<d<d_b$ this is no longer the case: the electronic wavefunctions
extend over both dots, and the $g_{\downarrow,\downarrow}$
pair correlation functions are very different depending
on the location of both electrons in the same or in different dots.
Note that transitions between different electronic
configurations vs. $d$ were recently identified theoretically also
for classical coupled dots \cite{peeters}:
We find that, for the small values of $N$ considered here,
the number of distinct phases and the spatial
distribution of electrons (as reflected in their
correlation functions) is drastically
modified by quantum effects.

The above findings are expected to be observable experimentally.
First, the calculated magnetic-field dependent addition spectra $A(N)$
present clear signatures of the phase transitions described above, as
illustrated in Fig.~4. Here $A(N)=E(N)-E(N-1)$ is obtained from
the many-body ground state energies of the N- and (N-1)-electron
systems, on the basis of single-particle states calculated
in the presence of the external magnetic field $B$.
The behaviour shown in the left, central, and right
panels is representative of the three phases.
Secondly, the changes in the magnetization induced
by one electron addition
---also accessible experimentally \cite{oosterkamp}---
are expected to follow a different pattern in each
phase (see the different sequence of quantum
numbers in Fig. 2).

A large experimental effort is presently devoted to
transport experiments in double QDs. In most cases,
the dots are obtained by gating a
two-dimensional electron gas 
(lateral confinement), and their coupling can be tuned
through a gate voltage
\cite{oosterkamp,waugh,livermore,wang,blick}.
Indeed, a set of
experiments has recently demonstrated a clear transition
to a `coherent' state with increasing coupling between dots
\cite{livermore,blick}. Outside this strong coupling regime,
however, experiments performed in the lateral geometry
have sofar evidenced classical interdot capacitance effects,
probably owing to the size of dots \cite{foot2}.
On the other hand, transport experiments are now available
on DQD structures with strong lateral confinement fabricated by
combined growth and etching techniques \cite{austing,wegscheider}.
The advantage is that
the number of electrons in the structure is limited, while
an accurate control on the inter-dot coupling is still possible
by designing samples with appropriate barrier thickness. Both aspects
are important to enhance many-body effects and to explore
intermediate coupling regimes. Recent experimental work on
these structures has focused on the weak-coupling
regimes \cite{austing}. We hope that further investigations will
be stimulated by the present work, since the relevant transitions are
now predicted quantitatively and should occur in
a range of parameters that is accessible to state-of-the-art experiments.
We expect that such studies will bring new insight into
electron-electron interaction effects in coupled
quantum nanostructures.

We acknowledge helpful discussions with C.~Calandra and G.~Goldoni.
This work was supported in part 
by INFM PRA99-SSQI, by the EC under the TMR 
Network ``Ultrafast Quantum Optoelectronics'', and by 
the MURST-40\% program ``Physics of Nanostructures''.
%\end{multicols}

\begin{figure}
%\centerline{\epsfig{file=fig1.ps,width=3.5in}}
\caption{\label{fig1}
Parameters entering the many-body Hamiltonian of a
double quantum dot with $\hbar \omega=$ 10 meV,
plotted as a function of the interdot distance $d$.
The hopping coefficient $t$
is shown together with two of the intra-dot and inter-dot
Coulomb integrals.
In the inset, the top panel shows the confinement
potential $V(z)$ (the barrier height is 200 meV)
and the corresponding symmetric and antisymmetric
single-particle wavefunctions,
$\left|s\right>$ and  $\left|a\right>$;
the bottom panel displays the localized states, 
$\left|1\right>$
and $\left|2\right>$, obtained as combinations
of $\left|s\right>$ and  $\left|a\right>$,
that are used as basis set for our calculation. }
\end{figure}

\begin{figure}
%\centerline{\epsfig{file=fig2.ps,width=3.5in}}
\caption{
\label{fig2} Energies of the ground state (thick line) and three
lowest excited states (thin lines) as a function of the inter-dot
distance, $d$, for a double QD occupied by $N$ electrons. For
$N=4$, the prominent single particle configurations contributing to
the many-body ground state are shown in the insets. Note that in
the intermediate phase a significant contribution comes also from
other Slater determinants (see text).  The $\pi$
molecular states are doubly degenerate because of the two-fold
degeneracy of the second shell in the single-dots ($p_x, p_y$).  }
\end{figure}

\begin{figure}
%\centerline{\epsfig{file=fig3.ps,width=3in}}
\caption{
\label{fig3}
Electronic pair correlation functions
for a double QD with $N=6$ electrons, calculated
at three values of the inter-dot distance, $d$,
corresponding to the different quantum phases.
The maps are plotted as a function of the difference
between the in-plane coordinates of the electrons in the pair
($x=x'-x''$ and $y=y'-y''$), while their $z$ coordinates
are fixed either at the center of the same dot
[$g_{\downarrow,\downarrow}(x,y,{z_1},{z_1})$; upper
panels] or at the center of different dots
[$g_{\downarrow,\downarrow}(x,y,{z_1},{z_2})$; lower
panels].
Here $\downarrow$ represents the minority spins.
}
\end{figure}

\begin{figure}
%\centerline{\epsfig{file=fig4.ps,width=3in}}
\caption{
\label{fig4}
Calculated addition spectra, $A(N)=E(N)-E(N-1)$, as a function
of magnetic field $B$ for a double QD, calculated at three values
of the inter-dot distance, $d$,
corresponding to the different quantum phases.
Labels indicate the ground state configuration of the $N$-electron
system. The energy zero is arbitrary.
}
\end{figure}

%\end{multicols}

\begin{thebibliography}{10}

\bibitem{reviews}
For reviews see: R.C. Ashoori, Nature {\bf 379}, 413 (1996);
L.P. Kouwenhoven et al., in {\it Mesoscopic Electron
Transport}, edited by L. Sohn et al. (Kluwer, Dordrecht, 1997);
and references therein.

\bibitem{hawrylak}
L. Jacak, P. Hawrylak, and A. W\'ojs,
{\it Quantum Dots} (Springer, Berlin, 1998).

\bibitem{perturb}
K.A. Matveev et al., Phys. Rev. B{\bf 53}, 1034 (1996);
J.M. Golden and B.I. Halperin, Phys. Rev. B{\bf 56}, 4716 (1996).

\bibitem{theory} Hubbard-like hamiltonians for QDs
are also used
---with different levels of approximations--- in
G.W. Bryant, Phys. Rev. B{\bf 48}, 8024 (1993);
G. Klimeck et al., Phys. Rev. B{\bf 50}, 2316 (1994);
J.J. Palacios et al., Phys. Rev. B{\bf 51}, 1769 (1995);
R. Kotlyar and S. Das Sarma, Phys. Rev. B{\bf 56}, 13235 (1997);
Y. Asano, Phys. Rev. B{\bf 58}, 1414 (1998);
Y. Tokura et al., Proc 24th Internat. Conference on the Physics of
Semiconductors, edited by D. Gershoni, World Scientific (1999),
in press.

\bibitem{austing}
D.G. Austing et al., Jpn. J. Appl. Phys. {\bf 36}, 1667 (1997).

\bibitem{tarucha}
S. Tarucha et al., Phys. Rev. Lett. {\bf 77}, 3613 (1996).

\bibitem{foot1} The same approach was introduced
for the case of isolated QDs giving addition spectra
in good agreement with experiments
[M.~Rontani, F.~Rossi, F.~Manghi, and E.~Molinari,
Appl.~Phys.~Lett.~{\bf 72,} 957 (1998); Phys.~Rev.~B {\bf 59,} 10165 (1999)].
In that case, however, the Hubbard hamiltonian reduces
to an effective one-body hamiltonian since $t$ and the inter-dot
integrals are zero.
\bibitem{notation}
Ground state configurations are labeled in the molecular
notation:
$\Sigma, \Pi, \Delta, \Phi, \Gamma, {\rm I}$ correspond to
$M=0, \ldots, 5$;
the left superscript is $2S+1$ ; $g$ and $u$ specify the symmetry
under inversion with respect to the center of the molecule;
$+$ and $-$ the symmetry under the $z \rightarrow -z$
reflection.
\bibitem{Eto} M. Eto,
Solid State Electronics {\bf 42} 1373 (1998)
\bibitem{peeters}
B. Partoens et al., Phys. Rev. Lett. {\bf 79}, 3990 (1997).
\bibitem{oosterkamp}
T.H. Oosterkamp et al., Phys. Rev. Lett. {\bf 80}, 4951 (1998).
\bibitem{waugh}
F.R. Waugh et al., Phys. Rev. Lett. {\bf 75}, 705 (1995).
\bibitem{livermore}
C. Livermore et al., Science {\bf 274}, 1332 (1996).
\bibitem{wang}
T.H. Wang and S. Tarucha, Appl. Phys. Lett. {\bf 71}, 2499 (1997).
\bibitem{blick}
R.H. Blick et al., Phys. Rev. Lett. {\bf 80}, 4032 (1998).

\bibitem{foot2} Interesting deviations from predictions of
simple capacitance models are however reported in
Refs.~\cite{oosterkamp,waugh}.

\bibitem{wegscheider} DQD with different geometries, obtained by
cleaved-edge or self-organized growth, have recently become
available but were sofar investigated mostly by optical experiments.
See e.g. G. Schedelbeck et al., Science {\bf 278}, 1792 (1997);
N.N. Ledentsov et al., Phys. Rev. B {\bf 54}, 8743 (1996);
R. Cingolani et al. (1999), unpublished.


\end{thebibliography}
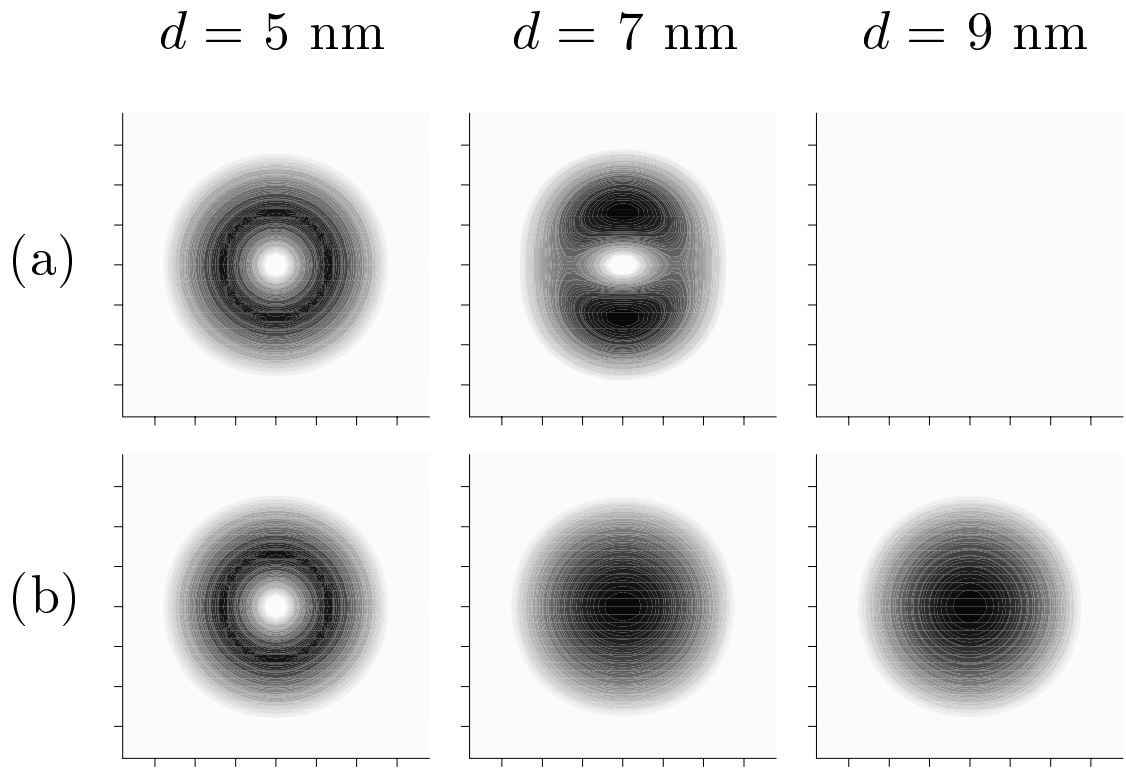
\end{document}